\documentclass[%
 reprint,
showpacs,
amsmath,amssymb,
aps,
prl,
]{revtex4-1}

\usepackage[usenames,dvipsnames,svgnames,table]{xcolor}
\usepackage{float}
\usepackage{graphicx}
\usepackage{dcolumn}
\usepackage{bm}

\usepackage[
margin=0.7in,
]{geometry}

\begin{document}

\preprint{APS/123-QED}

\title{Large electropositive cations as surfactants for the growth of
  polar epitaxial films}

\author{Alfred K. C. Cheung}
\author{Ilya Elfimov}%
\author{Mona Berciu}
\author{George A. Sawatzky}
\affiliation{Department of Physics and Astronomy, University of
  British Columbia, Vancouver, British Columbia, Canada V6T 1Z1} 


\date{\today}

\begin{abstract}
Using density functional theory (DFT) we demonstrate that the
adsorption of large cations such as potassium or cesium facilitates
the epitaxial growth of polar LaAlO$_{3}$ (LAO) on SrTiO$_{3}$
(STO). The low ionization potential of K favors efficient electron
transfer to the STO conduction band and results in a 2D electron gas
which exactly compensates for the diverging potential with increasing
layer thickness. For large cations like K or Cs, DFT total energy
considerations show that they remain adsorbed on the LAO surface and
do not enter substitutionally into LAO. These results suggest a novel
scheme for growing clean LAO/STO interface systems, and polar systems
in general, by performing the growth
process in the presence of large, low ionization potential alkali
metal ions.
\end{abstract}
\pacs{81.15.Aa, 73.20.-r, 68.35.B-, 73.20.Hb}

\maketitle


Since the discovery of the high mobility quasi-2-dimensional electron
gas (q-2DEG) at the $n$-type interface between LaAlO$_{3}$ (LAO) and
SrTiO$_{3}$ (STO) \cite{Ohtomo1}, enormous experimental and
theoretical efforts have been made to understand its origins and its
relation to the polar, crystallographic orientation induced electric
potential which diverges with film thickness
\cite{thiel1,nakagawa1,savoia1}. Although the issue
is still under debate, there is now a general consensus that
compensation of this diverging electric potential (the polar
catastrophe) is achieved through a combination of pure electronic
reconstruction \cite{hesper1,nakagawa1,savoia1}, interface defects
\cite{kalabukhov1,zunger1}, lattice distortions \cite{pickett1,vonk1},
and oxygen vacancies
\cite{levi1,eckstein1,cen1,siemons1,pauli1,pavlenko1,pavlenko2,zhang1,
  bristowe1,zhong1,chen1,li1,golden1,rusydi1,zunger1}. The last mechanism is
particularly noteworthy as recent studies \cite{rusydi1,zunger1} argue
that the formation of oxygen vacancies on the AlO$_{2}$ surface
provides a comprehensive explanation for the observed critical
thickness of four LAO layers required for the q-2DEG to appear.

The relative contributions of these mechanisms to removing the polar
catastrophe is expected to depend critically on the specific details
of the growth conditions. This explains why samples grown by different
groups can show widely dissimilar properties \cite{mannhart1}. The
important role played by defects and vacancies \cite{nakagawa1} also
explains why the growth of clean LAO/STO interfaces, necessary for
device applications, has proven to be such a challenge.

In this Letter we propose a solution for this challenge. We use
DFT calculations to show that a coverage
of 1/2 K per unit cell of the AlO$_2$ surface acts like a surfactant
that stabilizes the epitaxial growth of the polar LAO. Each K donates
an electron to the LAO/STO interface, generating a q-2DEG while 
keeping the top surface of LAO insulating. This q-2DEG fully compensates
for the polar catastrophe for any thickness of the LAO film, 
removing the reason for the appearance of defects, vacancies or
distortions. Moreover, because of their large size, the K ions cannot
enter substitutionally into LAO film.  As a result, after the
deposition of half a monolayer of K, the epitaxial LAO film will
grow cleanly under this surfactant. 

We emphasize that we use the term \textit{surfactant} as is
  customary in the crystal growth community
  \cite{tournie1,kandel1}. In surfactant-mediated epitaxy, the
  surfactant  facilitates the layer-by-layer growth
  of the film while always ``floating'' on its top surface
  \cite{tournie1,kaxiras1,mae1,cao1,kandel1}. An effective surfactant
  is energetically most favored on the film's
  surface \cite{kaxiras1}; when an adatom (in this case, La, Al,
  and O) arrives onto the surfactant surface, the adatom and a
  surfactant atom (in this case, K) exchange positions such that the
  surfactant re-emerges on the surface and the adatom is buried
  underneath; this process is then repeated
  \cite{tournie1,kandel1,kaxiras1,mae1,cao1,copel1}.

Our proposal is supported by past successes in using surfactants to stabilize
crystal surfaces
\cite{tournie1,kandel1,kaxiras1,mae1,cao1,copel1,heikens1,tixier1,young1}. As
a specific example, Heikens 
\textit{et al.} \cite{heikens1} stabilized (111)-terminated MnS,
otherwise impossible to grow due to the polar problem, by adsorption
of I$^{-}$ on the (111) surfaces. Another example is the growth of
GaN$_x$As$_{1-x}$ in the presence of Bi which was found
to stabilize the semiconductor-vacuum interface, increase surface
smoothness, and enhance nitrogen incorporation
\cite{tixier1,young1}. The use of alkali metals as electron donors is
also common. For instance, potassium deposition onto
YBa$_2$Cu$_3$O$_{6+x}$ was used for tuning its level of
doping \cite{fournier1,hossain1}.  We believe that, in general, the
use of low electron affinity surfactant layers of positive ions
can facilitate the growth of ionic materials with strongly polar
orientations.

\textit{Method}: All DFT calculations reported here are performed with
the Vienna \textit{ab initio} simulation package (VASP) \cite{kresse1}
using the projector augmented plane wave method
\cite{blochl1,kresse2}. The Perdew-Burke-Ernzerhof (PBE) functional
\cite{perdew1} is used for the exchange-correlation energy. The energy
cutoff for the plane-wave basis functions is 400 eV. For structural
optimization calculations, a $\Gamma$-centered (7,7,1) k-point mesh is
used. For density of states (DOS) calculations on the optimized
structures, a $\Gamma$-centered (17,17,1) k-point mesh is used. 

\begin{figure}
\includegraphics[width=\columnwidth]{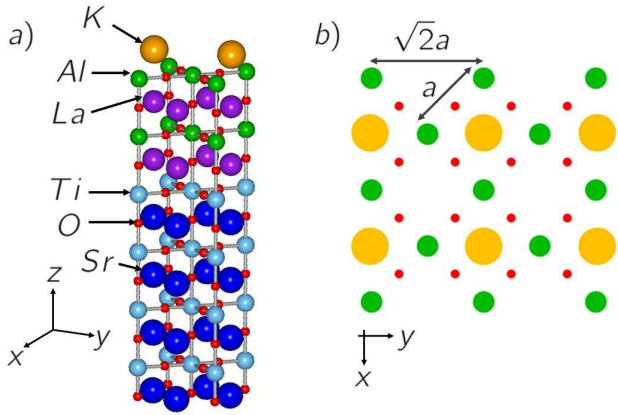}
\caption{(Color online) (a) K$_{ads}$(LAO)$_{m}$(STO)$_{4}$ for $m=2$
  (initial structure). K atoms are placed on the surface AlO$_{2}$
  layer in the middle of the squares formed by oxygen atoms. This is
  illustrated in panel (b) which shows a top view of the K adsorption
  sites on the surface AlO$_{2}$ layer. The simulated square unit cell
  has a side length of $\sqrt{2}a$, where $a=3.913$ \AA{} is the
  theoretical lattice constant of bulk STO obtained from DFT using the
  PBE functional.}
\label{fig:1}
\end{figure}

The structures simulated are comprised of $m$ unit layers of LAO on
top of 4 unit layers of STO substrate. The interface is $n$-type
(TiO$_2$/LaO). K atoms are adsorbed on the surface AlO$_{2}$ layer of
LAO at a concentration of 1 adsorbed atom per 2 lateral AlO$_{2}$ unit
cells, {\em i.e.} the theoretical concentration needed for exact compensation of
the polar problem (other coverages are considered in the Supplemental Material
\cite{supp}). We refer to this general structure as
K$_{ads}$(LAO)$_{m}$(STO)$_{4}$. The structure for $m=2$ is shown in
Fig.~\ref{fig:1}a. We choose to place the K atoms above the center of
squares formed by the oxygen atoms in the surface AlO$_{2}$ layer, as
shown in Fig.~\ref{fig:1}b. This position should be the most
stable configuration because upon donating its $4s$ electron, a K$^{+}$
cation should be attracted to each O$^{2-}$ anion.  Indeed, alternate
positions were found 
to be unstable towards relaxing back into this chosen position, thus
validating our choice.

The lateral lattice constant is fixed at $\sqrt{2}a$ where $a=3.913$
\AA{} is the calculated  DFT-PBE lattice constant for bulk
STO. Atoms of the bottom-most STO unit layer are kept fixed so as to
simulate the effect of the  infinitely thick substrate. All
other atoms are allowed to relax 
along the $z$-direction until the force on each atom is less than 0.02
eV/\AA{}. 15 \AA{} of vacuum is put on top of each slab to minimize
interactions between periodic copies of the slab. Dipole corrections
to the total energy and electric potential are used to remove any
remaining spurious contributions due to periodic boundary conditions
\cite{makov1}. 

\begin{figure}[!h]
\includegraphics[width=\columnwidth]{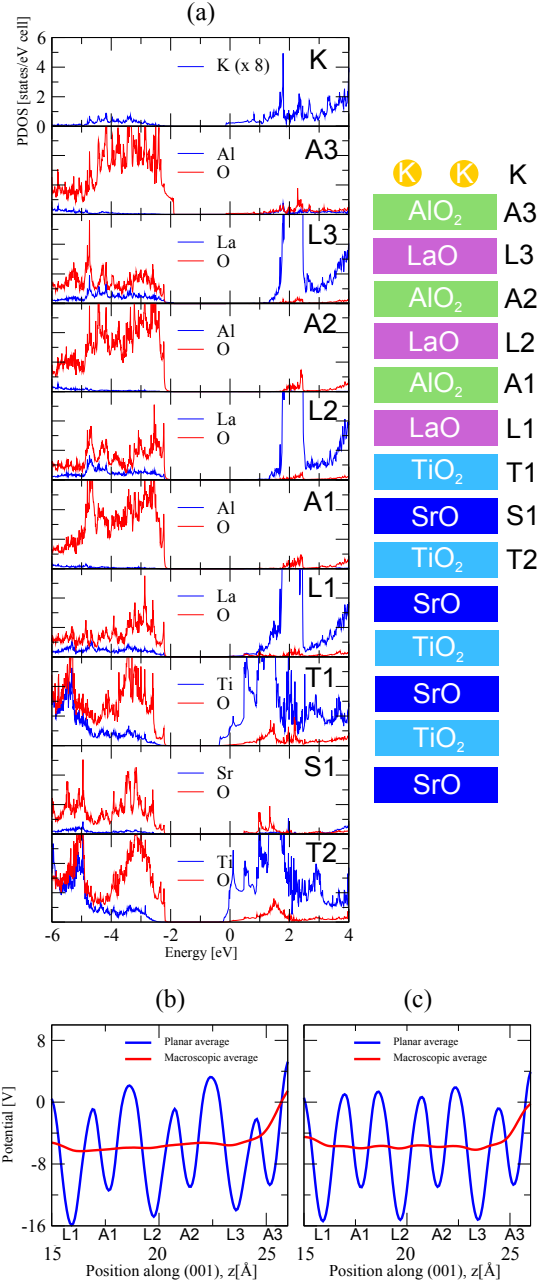}
\caption{(Color online) (a) Layer and element projected DOS for
  K$_{ads}$(LAO)$_{3}$(STO)$_{4}$. The Fermi energy is at 0. The scale
  is the same for all panels. Upon 
  adsorption of K, the TiO$_2$ layers become conducting. This is
  due to the donation of the K $4s$ electron to the Ti $3d$
  conduction bands. The conduction electron density is greatest for
  the TiO$_2$ layer closest to the interface. The planar-average and macroscopic average
  electric potentials (as defined in Eq.~\ref{eq:-1} and Eq.~\ref{eq:0}) are plotted versus the $z$-position along
 the (001) direction of the supercell, within the LAO region, for the 
 system (b) without and (c) with K. On the scale of the plot, no residual potential is observed for the K adsorbed system. 
	In contrast, a potential buildup of the order of 2 eV exists for
	the bare system without K.} 
\label{fig:2}
\end{figure}

\textit{Compensation by potassium electron donation}: Layer and
element projected partial densities of states (PDOS) for the case of
$m=3$ are shown in Fig.~\ref{fig:2}a. Upon adsorption of K, electrons
are donated into the Ti $3d$ conduction bands. The conduction electron
density is greatest for the titanate layer closest to the interface
and decays for layers further away, forming a q-2DEG as in pure
electronic reconstruction. In both cases, the extent of the electron
transfer is limited by an associated energy cost. For pure electronic
reconstruction, this penalty is the band gap between the valence band
of LAO and the conduction band of STO. In the present case, this is
given by the binding energy of the $4s$ electron of the adsorbed
K. This parameter controls how much of the diverging potential across
LAO is compensated. To evaluate their efficiency, we look for residual
potential buildup across LAO along the $z$-direction. If $V(x,y,z)$ is the electric potential function, 
then the planar-average potential along $z$, $\bar{V}(z)$, is defined by:
\begin{equation}
\bar{V}(z)=\frac{1}{S}\int_S dx dy V(x,y,z),
\label{eq:-1}
\end{equation}
where $S$ is area of the lateral unit cell. It is also useful to define the 
macroscopic average potential, $\bar{\bar{V}}(z)$ \cite{peressi1}:
\begin{equation}
\bar{\bar{V}}(z)=\frac{1}{a_z}\int_{z-a_z/2}^{z+a_z/2} dz^\prime \bar{V}(z^\prime).
\label{eq:0}
\end{equation}
Here, $a_z$ refers to the lattice constant of LAO in the $z$-direction. In other words, $\bar{\bar{V}}(z)$ averages out oscillations within one unit cell. Because of relaxation, especially in the uncompensated film, a constant $a_z$ is actually ill-defined. For convenience, we take as $a_z$ the relaxed thickness of LAO divided by the number of unit cells. One can also generalize the notion of the macroscopic average to account for the interface with STO \cite{peressi1}, but since we are only interested in the potential buildup across LAO, we use the definition in Eq.~\ref{eq:0}.

In Fig.~\ref{fig:2}b, we plot $\bar{V}(z)$ and $\bar{\bar{V}}(z)$
within the LAO region for the system without adsorbed K. The same is shown for the
system with adsorbed K in Fig.~\ref{fig:2}c. No potential buildup is observed 
across LAO in the system with K, whereas a potential buildup of $\sim$ 2 eV is
present in the system without K. This is evident looking at the overall slopes in 
the macroscopic potentials, as well as the positions of the minima in the planar-average potentials.

Ref.~\cite{arras1} proposed a different way to
  remove this potential buildup, by placing metallic overlayers
  {\it after growth}; this has since been done with Co in
  Ref. \cite{lesne1}. However, in this case the top
  layer is also metallic and separating its conductivity from that of
  the q-2DEG is difficult. This is not an issue for our
  proposal, where the K surfactant layer at a concentration of 1/2 per 
	unit cell remains insulating since it has donated its charge carriers 
	to the interface. Furthermore, since the metal capping is done after
  growth,  the tendency for defects and vacancies to form during
  growth is not mitigated. In contrast, as we argue below, the crystal
  grows underneath the K surfactant layer free of such defects.

It has been shown that in the absence of oxygen vacancies or other
defects, the onset of electronic reconstruction is delayed through
formation of polar distortions within the LaO layers \cite{pickett1},
which create internal compensating dipoles that screen the electric
potential. Below a critical thickness of 4-5 unit LAO layers, this
suffices to partially compensate the polar potential and electronic
reconstruction does not occur. In thicker films with larger potential
buildup, the compensation 
requires electronic reconstruction, which in turn removes the need for
these polar distortions within the LaO layers.  It is thus worthwhile
to consider whether the donation of the K $4s$ electron to Ti $3d$
bands occurs for all thicknesses of LAO, or whether a critical
thickness exists for similar reasons.

The total DOS for $1\le m\le 6$, with and without K adsorption, are
shown in Fig.~\ref{fig:3} (projected densities of states are shown in
the Supplemental Material \cite{supp}). The critical thickness is defined to
be the number of layers of LAO at which the system without K becomes
conducting. From our calculations this is at 4 unit layers of LAO
(Fig.~\ref{fig:3}c). The system with adsorbed K is found to be
conducting for all thicknesses simulated: even at one unit layer of
LAO, where the potential across the film is smallest, K
gives up its $4s$ electron to the Ti $3d$ bands. As 
expected, the polar distortions within LAO are eliminated upon adsorption of K.

\begin{figure}
\includegraphics[width=\columnwidth]{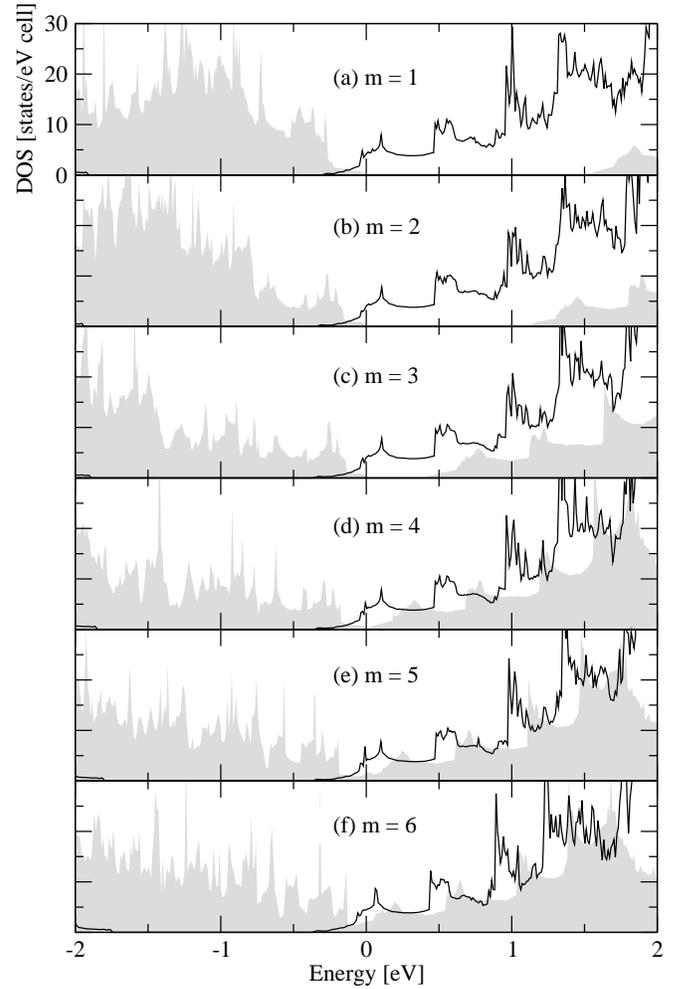}
\caption{Total DOS for (LAO)$_m$(STO)$_4$, $m=1,2,3,4,5,$ and $6$, with
  (black line) and without (shaded gray) K adsorption. The Fermi
  energy is at 0. The scale is the same for all panels. The critical
  thickness at which the system without K 
  undergoes electronic reconstruction and becomes conducting is 4 unit
  layers of LAO. In contrast, the system with K is conducting at all
  LAO thicknesses simulated.} 
\label{fig:3}
\end{figure}

\begin{figure}
\includegraphics[width=\columnwidth, clip]{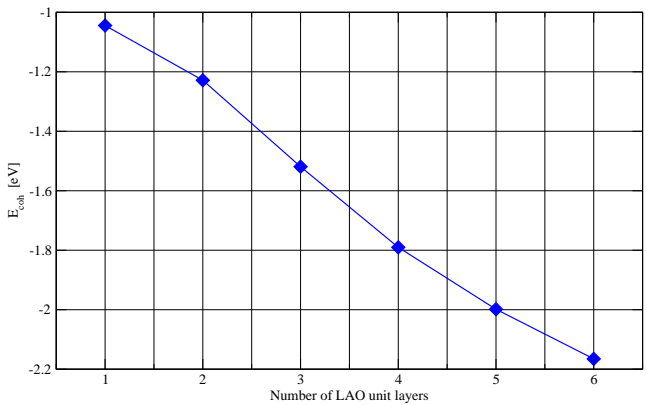}
\caption{(Color online) Cohesive energy per adsorbed K, defined in
  Eq. (\ref{eq:1}),  as a function
  of the thickness of LAO. The system 
  with K becomes more stable as the thickness increases,
  reflecting the increasing electric potential across LAO as LAO
  thickness is increased in the bare system.  
  The cohesive energy is approximately 1 eV for 1 unit layer of LAO, 
	increasing to more than 2 eV as LAO thickness is increased, 
	a significant energy reduction.} 
\label{fig:4}
\end{figure}

\textit{Cohesive energies}: Figs.~\ref{fig:2} and~\ref{fig:3} show
that K adsorption negates the polar catastrophe and stabilizes the
LAO/STO heterojunction. The additional energetic stability can be
quantified by calculating the cohesive energy of the adsorbed system
with respect to the system without K and in the absence of other
defects. We define the LAO thickness dependent cohesive energy
$E_{coh}(m)$ as: 
\begin{equation}
E_{coh}(m)=E_{K_{ads}LAO_{m}STO_{4}}-E_{LAO_{m}STO_{4}}-E_{K},
\label{eq:1}
\end{equation}
where $m$ is the number of unit layers of LAO,
$E_{K_{ads}LAO_{m}STO_{4}}$ is the total energy of the system with K
adsorption, $E_{LAO_{m}STO_{4}}$ is the total energy of the system
without K adsorption (using the same size of lateral unit cell), and
$E_K$ is the energy per atom of K crystal (body-centered cubic). Hence, $E_{coh}$ can be
interpreted as the cohesive energy per adsorbed K, with respect to
metallic K. It is plotted as a 
function of $m$ in Fig.~\ref{fig:4}.  

Two observations can be made: (1) The K-adsorbed system becomes more
stable relative to the original system as the number of LAO
unit layers is increased. This is a trivial consequence of the
increasing potential buildup across LAO as more LAO layers are
stacked---the greater the potential buildup, the more energetically
unstable the original system becomes, and the greater the energy
reduction when the potential difference is eliminated by electron
transfer from adsorbed K; (2) For the thicknesses considered,
$|E_{coh}|$ ranges from 1 eV to more than 2 eV (1 eV $\sim 12000$ degrees Kelvin), a
very substantial energetic stabilization.  

\textit{Undesired substitutions of K into LAO}: Our discussion above
argues for K surfactant adsorption as an effective mechanism for removing the
polar catastrophe in LAO/STO(001) to allow for the growth of clean
heterostructures. This proposal hinges on K remaining on the AlO$_{2}$ surface
instead of entering the bulk of the LAO by substitution of La or Al
ions. This is highly unlikely given the large
differences in ionic radii between K$^{+}$ (151 pm), and La$^{3+}$
(116 pm) and Al$^{3+}$ (54 pm) \cite{CRC}. We confirm this
by calculating the total energy difference between systems where K
exchanges position with an La or Al ion close to the surface, and the
system with K adsorbed on the surface. Calculations are done
for $m=5$ and are expected to  be representative for all $m$.

Consider first the exchange of K with a La ion in the LaO layer
closest to the surface. Two substitutions are possible: (1) K
exchanges position with the La beneath it, or (2) with the
other La  (\textit{cf.} Fig.~\ref{fig:1}). The corresponding energy
cost per substituted K is 1.93 eV for the first
case, and 2.27 eV for the second one. For the
exchange of K with an Al ion in the surface layer, there is only one
option possible. In a striking display of how energetically
unfavorable this substitution is, our relaxation of the substituted
structure resulted in the potassium cation pushing its way back above
the rest of the structure. Substitution into deeper layers of
LaO/AlO$_{2}$ is expected to be just as, if not more, energetically
costly. This proves that the energy cost for K substitution of La/Al
within the bulk of LAO is very large, confirming our hypothesis.

To summarize, we have shown that K adsorbed
on the AlO$_2$ surface of an $n$-type LAO/STO(001) heterojunction compensates the
diverging electric potential across LAO by donating its $4s$ electron
to the Ti $3d$ conduction band. The electron transfer is found
to occur at all thicknesses of LAO studied, below and above the
critical thickness for electronic reconstruction in the system without
K. Furthermore, the compensation is highly efficient, with very little
residual potential buildup across LAO. The cohesive energy is
calculated to be approximately 1 eV per adsorbed K for 1 unit layer of LAO, 
increasing to more than 2 eV as LAO thickness is increased, proving  how
energetically favorable K adsorption is. Finally, substitution of K into
layers of LAO by exchange with La or Al is demonstrated to be extremely unfavorable.  

Taken together, these results suggest an elegant scheme for growing
clean LAO/STO heterojunctions. By executing the growth process in the
presence of alkali metals with low ionization energies, the diverging
potential is eliminated without appealing to the myriad of other --
often uncontrollable -- compensation mechanisms. On a practical level,
this requires only a very small surface concentration of K on the
order of 10$^{14}$/cm$^{2}$. This can be obtained with an exposure to
K after deposition of very few layers with the O source closed, and then removing the K
source.  This could also solve the issue of
  oxidation of K, if it occurs \cite{note}. 
As we show in the Supplemental Material \cite{supp}, the
cohesive energy for K adsorbed beyond the 1/2 per unit cell (ideal)
concentration is significantly smaller than for K absorbed at the
ideal concentration.  Hence, if the substrate temperature is
sufficiently high, any extra K will evaporate, preventing the
formation of an undesired thick metallic overlayer of K on the
surface. Their large size also prevents the 
  alkali metal cations from being incorporated into the film during
  growth, guaranteeing their role as surfactants. Smaller
  $3d$ transition metal ions (TM) would not be suitable because most
  of them form LaTMO$_3$ perovskite structures with similar lattice
  constants as LAO so it is reasonable to expect that they would
  substitute for Al during growth.

More importantly, our proposed scheme can become a new paradigm for
growing LAO/STO(001) interfaces wherein the resulting samples will no
longer have properties highly sensitive to growth conditions. The
problem of electronic and structural properties being attributed to
different defects in samples prepared differently has been identified
as a major potential pitfall for the development of applications based
on the LAO/STO interface \cite{mannhart1}. We believe that the
method we have proposed here is a solution to this problem, not just
for LAO/STO, but for many systems in which clean growth in a
particular direction is impeded by the polar catastrophe.


\begin{thebibliography}{44}
\bibitem{note} In principle, it is possible that  K may undergo fast
    oxidation with the resulting oxide accumulating on the surface,
    impeding further growth. Whether this really happens can
    only be ascertained through   experiment. If it does, a potential
    solution  is to grow the first two  or three layers without K,
    allowing 
    the internal buckling to  compensate for the potential buildup
    \cite{pickett1}. The  O, La, and Al sources are then shut off,
    after which
    K is absorbed onto the exposed AlO$_2$ surface. This will  remove
    the buckling since the K will take over in compensating for the
    polar 
problem. The K source is then turned off and the growth
of the LAO is restarted, with the film growing cleanly underneath the
layer of K surfactant.
\end{thebibliography}
\end{document}